\documentclass[aps,prl, preprintnumbers,groupedaddress,nofootinbib]{revtex4}
\usepackage{graphicx}
\usepackage{latexsym}
\usepackage{amsfonts}
\usepackage{amssymb}
\usepackage{amsmath}
\usepackage{slashed}
\usepackage{feynmp}
\usepackage{comment}
\def\be{\begin{equation}}
\def\ee{\end{equation}}
\def\bea{\begin{eqnarray}}
\def\eea{\end{eqnarray}}

\def\ltap{\ \raise.3ex\hbox{$<$\kern-.75em\lower1ex\hbox{$\sim$}}\ }
\def\gtap{\ \raise.3ex\hbox{$>$\kern-.75em\lower1ex\hbox{$\sim$}}\ }

\newcommand{\vev}[1]{ \left\langle {#1} \right\rangle }

\newcommand{\Mev}{{\rm MeV}}
\newcommand{\gev}{{\rm GeV}}

\newcommand{\mev}{{\rm MeV}}

\pdfoutput=1
\begin{document}

\title{
Dark Matter Self-Interactions and Light Force Carriers
}

\author{Matthew R.~Buckley$^1$ and Patrick J.~Fox$^2$}
\affiliation{$^1$Department of Physics, California Institute of Technology, Pasadena, CA 91125, USA}
\affiliation{$^2$Fermi National Accelerator Laboratory, Batavia, IL 60510, USA}
\preprint{CALT-68-2758}
\preprint{FERMILAB-PUB-09-560-T}
\date{\today}

\begin{abstract}
Recent observations from PAMELA, FERMI, and ATIC point to a new source of high energy cosmic rays.  If these signals are due to annihilating dark matter, the annihilation cross section in the present day must be substantially larger than that necessary for thermal freeze-out in the early universe.  A new force, mediated by a particle of mass $\mathcal{O}(100~\mbox{MeV})$, leading to a velocity dependent annihilation cross section - a `Sommerfeld enhancement' - has been proposed as a possible explanation.  We point out that such models necessarily increase the dark matter (DM) self-scattering cross section, and use observational bounds on the amount of DM-DM scattering allowed in various astrophysical systems to place constraints on the mass and couplings of the light mediator.
\end{abstract}
\maketitle

\section{Introduction}

It is well known that the Standard Model (SM) particles are not the primary constituents of the Universe. Dark energy makes up $72\%$ of the energy budget, while dark matter makes up $24\%$, and only the remaining small fraction is due to the SM~\cite{Spergel:2006hy}.
The nature of dark matter (DM) is one of the major puzzles facing physics today: other than its energy density and gravitational interactions, little about it is known with any certainty. 

Many solutions to the hierarchy problem naturally contain new stable massive particles.  Their masses are typically at the weak scale and their annihilation cross section is of the right order of magnitude such that, if they were thermally produced in the early universe, they would have the correct abundance today to be the DM.  For this so-called ``WIMP miracle'' (Weakly Interacting Massive Particle) to occur, and the DM to be a thermal relic, the annihilation cross section at the time of freeze-out, when the temperature was $T\sim m_\chi/25$, 
must have been $\langle \sigma v\rangle \sim 3 \times 10^{-26}$~cm$^3/$s.  It is usually assumed that this annihilation cross section is velocity independent and that in the present epoch the DM is annihilating with the same cross section. This is an appealing possibility, but by no means the only one.

Recently there have been a series of observations of high energy electrons, positrons, and gamma rays from HEAT \cite{Barwick:1997ig}, AMS-01 \cite{Aguilar:2007yf}, PPB-BETS \cite{Torii:2008xu}, PAMELA \cite{Adriani:2008zr}, ATIC \cite{Chang:2008zzr}, and the Fermi Gamma Ray Space Telescope (FGST) \cite{Abdo:2009zk}.  These measurements are seemingly at odds with what is predicted from the secondary production of anti-particles from cosmic-ray propagation.  These excesses may be explained by astrophysical processes -- for instance nearby pulsars~\cite{Hooper:2008kg,Profumo:2008ms} or supernovae remnants~\cite{Blasi:2009bd} may be a source for high energy positrons and electrons -- or they may be due to DM annihilating in our galaxy.

If these anomalies are related to DM annihilation \cite{Sommerfeld1,Baltz:2001ir,Barger:2008su,Cholis:2008vb,Bergstrom:2009fa,Harnik:2008uu,Nelson:2008hj,Cholis:2008qq,Zurek:2008qg,Fox:2008kb,Chen:2008dh,Cholis:2008wq,Hooper:2009fj,Hooper:2009gm,Meade:2009iu,Cirelli:2008jk}, the necessary value of $\langle \sigma v \rangle$ in the present day must be larger than the thermal freeze-out value of $\langle \sigma v\rangle \sim 3 \times 10^{-26}$~cm$^3/$s by a factor of approximately $10^{2-3}$ (see {\it e.g.}~\cite{Cholis:2008hb}). It is possible that this enhancement is purely structural in nature: if sufficient halo substructure exists wherein the DM density, $\rho$, is larger than that expected from Galactic simulations, the annihilation rate, which scales as $\rho^2$, would be increased. Alternatively, the presence of a new force interacting with DM and mediated by a light boson (scalar or vector) could provide a velocity-dependent annihilation cross section via the Sommerfeld effect \cite{Sommerfeld1,Cirelli:2008pk,ArkaniHamed:2008qn,Kuhlen:2009kx}. This enhancement typically scales as $v^{-1}$. However, due to these new forces the DM particles $\chi$ can develop bound states, near which the enhancement is much larger, proportional to $\sim v^{-2}$.

For the remainder of this paper, we shall take the Sommerfeld enhancement via some massive (but relatively light) boson $\phi$ as the explanation for the cosmic ray data. We then are interested in the associated enhancement of the {\it self-scattering} cross section of dark matter ({\it i.e.}~$\chi$-$\chi$ or $\chi$-$\bar{\chi}$ scattering, not $\chi$-nucleon scattering). The same non-perturbative effect that alters the wavefunction at zero distance will also cause a phase shift at infinity. As a result, an increase in the scattering cross section goes hand in hand with the Sommerfeld boost to the annihilation rate. However, as we shall see, though both these boosts occur at the same points in parameter space, the numerical size of the boost differs greatly between scattering and annihilation.

It has been shown, both by $N$-body simulation and direct observation of the system of colliding galaxy clusters known as the Bullet cluster, that dark matter must be, to good approximation, collisionless. From these observations, we may place limits on the mediator mass and coupling of any new force at work in the dark sector. Constraints from dark matter collisions in the case of a massless mediator have been considered previously \cite{Ackerman:2008gi,Feng:2009mn}.  

In the remainder of this paper we first derive a general expression for the scattering cross section of two dark matter particles interacting via some light force carrier.  This expression must be solved numerically, so we also find an approximate formula for the cross section, which is valid in the regime of very small mediator mass. After this, we consider the various bounds on the self-scattering cross section and the characteristics of the systems from which these bounds are extracted. Using the approximate parameters of these systems we then constrain the coupling and mass of possible new dark forces.


\section{Scattering and Annihilation}

\begin{figure}[t]
\includegraphics[width=0.75\columnwidth]{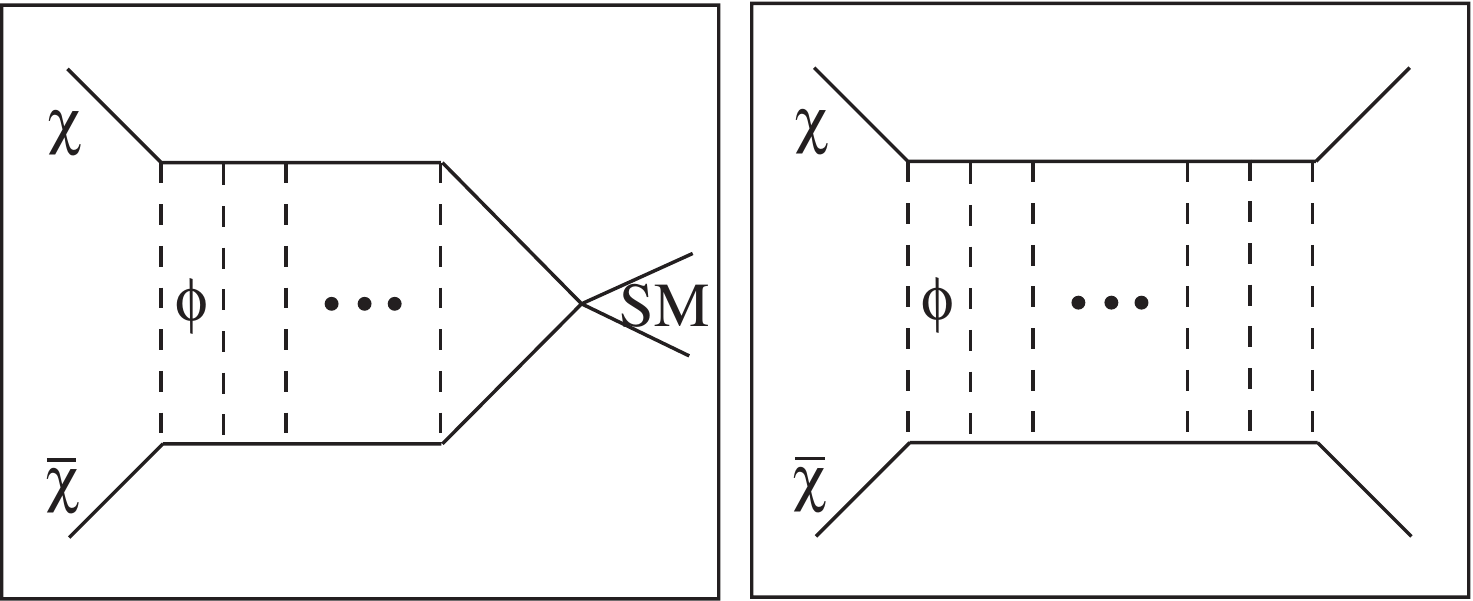}
\caption{Non-perturbative ``ladder'' diagrams corresponding to the formation of a $\chi-\bar{\chi}$ bound state for annihilation (left) and scattering (right). \label{fig:ladder}}
\end{figure}

We wish to consider the interactions between two slowly moving DM particles exchanging light force carriers. As the DM particles are non-relativistic they may exchange multiple bosons while undergoing either an annihilation or scattering process, as shown in Fig.~\ref{fig:ladder}.  This is a non-perturbative effect that must be resummed, which is done by solving the Schr\"odinger equation for the reduced system.  The potential between the two particles alters the wavefunction of the reduced system both at $r=0$, affecting the annihilation rate, and at $r\rightarrow \infty$, affecting the scattering cross section.  Both scattering and annihilation involve similar diagrams, the only difference between the two is an insertion of the short distance operator responsible for DM annihilation.  Thus, whenever the parameters of the system are such that there is a large enhancement in the annihilation cross section, then the self-scattering cross section is also enhanced.  Since the annihilation diagram must end in the $\chi-\bar{\chi}$ self-annihilation interaction which is not present in the scattering, there is no reason to expect that the size of both enhancements will be equivalent.

We consider two DM particles of mass $m_\chi$, with the force between them given by a potential $V(r)$. For the rest of this article we will restrict ourselves to the case of a Yukawa potential, generated by the exchange of a boson of mass $m_\phi$ which couples to the DM with coupling strength $\lambda$:
\be
V(r) = -\frac{\alpha}{r} e^{-m_\phi r} \label{eq:potential}~.
\ee
Here $\alpha = \lambda^2/4\pi$.  The wavefunction of the reduced system is $\psi(r)=\sum_{\ell,m} R_{\ell}(r) Y_{\ell,m}(\theta,\phi)$, where the radial wavefunction, $R_\ell(r)$, satisfies the radial Schr\"odinger equation,
\be
\frac{1}{r^2}\frac{d}{dr}\left(r^2 \frac{dR_\ell}{dr}\right) +\left[k^2 - \frac{\ell(\ell+1)}{r^2}-2\mu V(r)\right]R_\ell = 0~. \label{eq:radialschrodinger}
\ee
Here, $\mu=m_\chi/2$ is the reduced mass and $k=\mu v_{\rm rel}$ is the momentum in the reduced system. It is useful to introduce $\chi_\ell \equiv r R_\ell$ and $x \equiv m_\chi \alpha r$, in terms of which Eq.~(\ref{eq:radialschrodinger}) becomes
\be
\chi_\ell^{\prime\prime}+\left[\left(\frac{v_{\rm rel}}{2\alpha}\right)^2-\frac{\ell(\ell+1)}{x^2}+\frac{1}{x}e^{-\frac{m_\phi x}{\alpha m_\chi}}\right]\chi_\ell=0~.\label{eq:rescaledschrodinger}
\ee
We solve this equation with the boundary conditions that $\chi$ is regular at the origin and at large $r$ behaves as
\be
\chi_\ell \rightarrow \alpha\, m_\chi \sin \left(\frac{v_{\rm rel}}{2\alpha} x -\frac{\pi \ell}{2} +\delta_\ell \right)~.
\label{eq:asymptote}
\ee
The differential scattering cross section is given by
\be
\frac{d\sigma}{d\Omega} = \frac{1}{k^2}\left| \sum_\ell (2\ell +1) e^{i\delta_\ell}P_\ell (\cos\theta) \sin\delta_\ell 
\right|^2~,
\label{eq:diffcrosssection}
\ee
and the total scattering cross section is then given by the sum over all angular momenta $\ell$:
\begin{equation}
\sigma = \frac{4\pi}{k^2}\sum_{\ell=0}^\infty (2\ell +1) \sin^2 \delta_\ell~. 
\label{eq:sigmadef}
\end{equation}
A quantity of interest when discussing observational constraints is the transfer cross section, $\sigma_{\rm tr}$, which is a weighted average of the differential cross section
\bea
\sigma_{\rm tr} &\equiv& \int d\Omega (1-\cos\theta) \frac{d\sigma}{d\Omega} \nonumber \\
&=& \frac{4\pi}{k^2}\sum_{\ell=0}^{\infty} \left[  (2\ell +1)\sin^2\delta_\ell-2(\ell+1) \sin \delta_\ell \sin\delta_{\ell+1} \cos(\delta_{\ell+1}-\delta_\ell)\right]~.
\label{eq:sigmatr}
\eea
This weighted cross section controls the rate at which energy is transferred between particles in a collision. 

For the case of annihilation the enhancement is determined from the value of the wavefunction at the origin, $|\psi(0)|^2$.  At low velocities the attractive potential distorts the wavefunction, increasing it at the origin -- the Sommerfeld enhancement~\cite{Sommerfeld1}.  For a Yukawa potential this enhancement at low velocities scales as $\sim 1/v$ \cite{Lattanzi:2008qa,MarchRussell:2008tu,ArkaniHamed:2008qn} but saturates at velocities of order $v \sim \sqrt{\alpha m_\phi/m_\chi}$.  For some low velocities there are particular points in parameter space where there is a light resonance due to a bound state in the potential \cite{Cassel:2009wt,Iengo:2009xf,Iengo:2009ni}, which can greatly increase the annihilation cross section.  These same bound states will also lead to a large enhancement in the scattering cross-section.
  
As the Yukawa potential does not have analytical solutions, we are required to numerically solve Eq.~(\ref{eq:rescaledschrodinger}). Hence, we cannot sum over all possible values of the orbital angular momentum $\ell$ to find the total cross section. It is therefore useful to determine what the largest value of $\ell$ that is relevant to the sum in Eqs.~(\ref{eq:sigmadef}) and (\ref{eq:sigmatr}).  We can estimate the relevant angular momentum of the system by considering the momentum $k$, and impact parameter $b$, of the incoming particle.  We would expect $L$, the largest angular momentum needed to describe the interaction, to be $L \equiv k b_{\rm max}= \mu v_{\rm rel} b_{\rm max}$.  We estimate $b_{\rm max}$, the largest impact parameter relevant to the scattering, to be the separation at which the potential energy is comparable to the kinetic energy.  The potential is finite range, so for $r\gg b_{\rm max}$ there is little scattering and the contribution of the corresponding $\ell$-modes to the total cross section should be small.

\begin{figure}[ht]
\centerline{
\includegraphics[width=0.4\columnwidth]{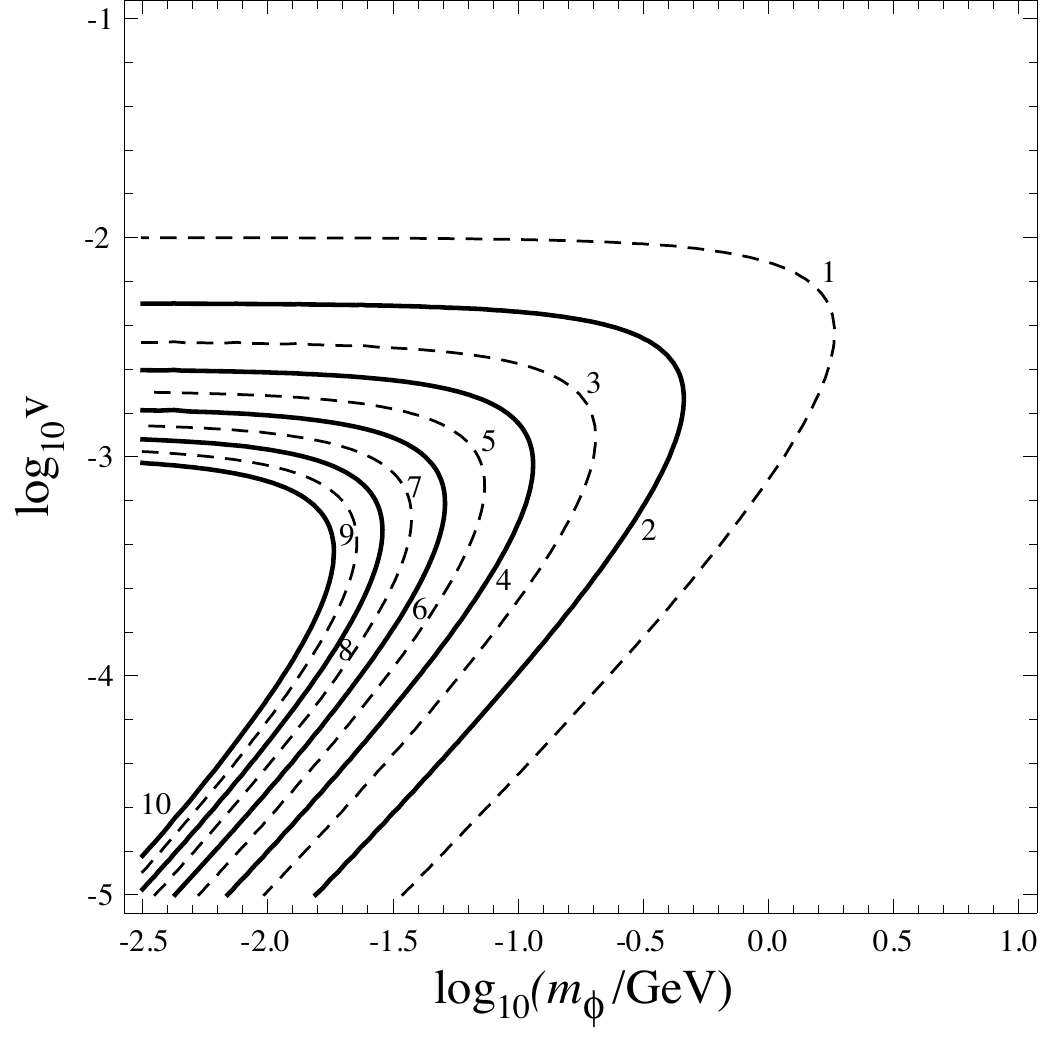}\hspace{0.07\columnwidth}
\includegraphics[width=0.4\columnwidth]{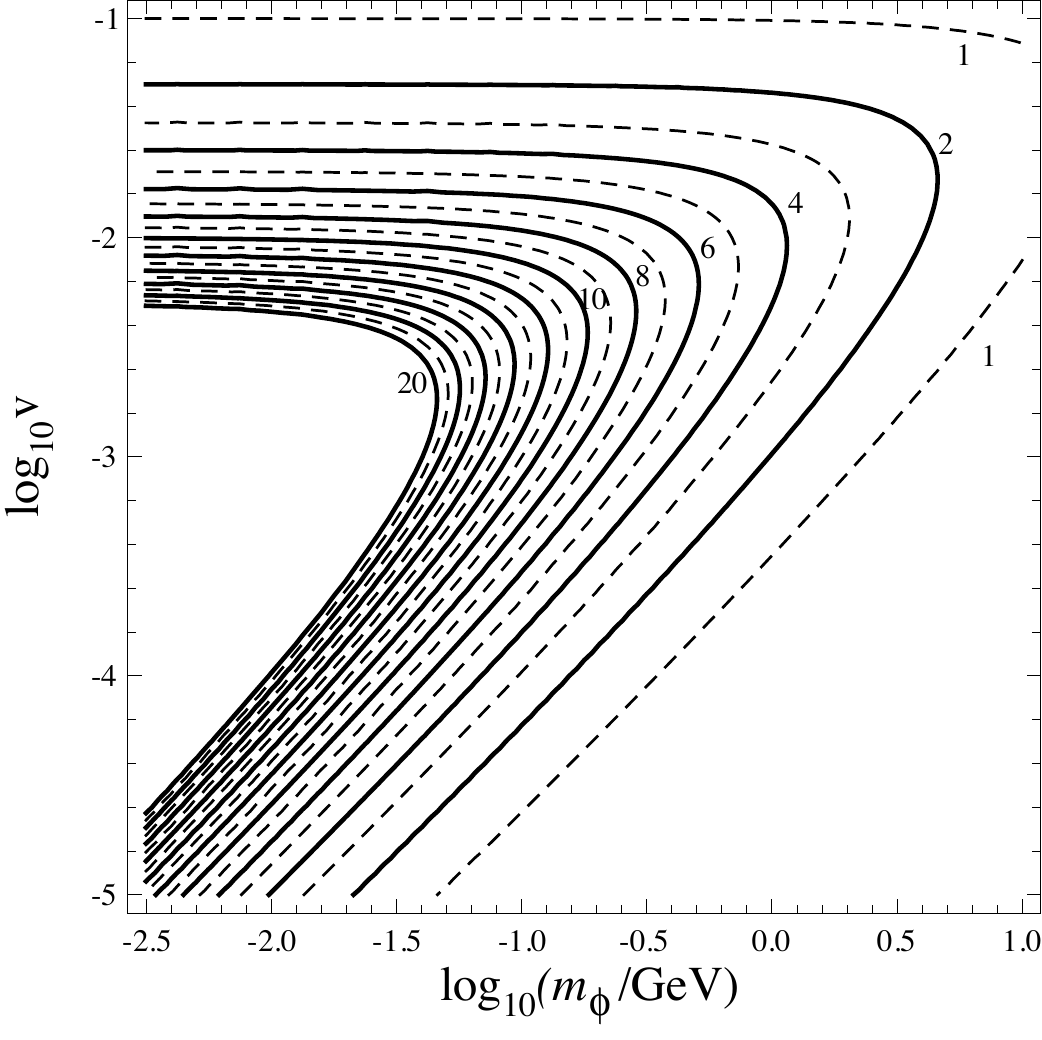}
}
\caption{The maximum angular momentum $L = m_\chi v b_{\rm max}$ as a function of the mediator mass $m_\phi$ for DM of velocity $v$ colliding head-on. In the left-hand plot we have chosen $\alpha = 0.01$ and in the right-hand plot $\alpha = 0.1$. In both cases $m_\chi=500\,\gev$. Each contour line corresponds to an increase of $L$ by one from the previous. \label{fig:lmax}}
\end{figure}

For values of $m_\phi$ large relative to the DM's kinetic energy, it is clear that $b_{\rm max} \ltap m_\phi^{-1}$, since the potential is controlled by the exponential.  In general the largest impact relevant impact parameter is given by the solution to the equation,
\begin{equation}
\frac{1}{2}\mu v_{\rm rel}^2 = \frac{\alpha}{b_{\rm max}} e^{-m_\phi b_{\rm max}}~. \label{eq:implicit}
\end{equation}
The resulting values of $L$ are shown in Fig.~\ref{fig:lmax} for two values of the coupling: $\alpha = 0.1$ and $0.01$.

Before solving for the phase shifts numerically it is useful to consider a simple approximation of the cross section in a Yukawa potential.  We make the approximation that for all $\ell\leq L$ the phase shift is maximal {\it i.e.} $\sin \delta_\ell \sim1$ and that for all $\ell > L$, the phase shift is zero.  This approximation is borne out in numerical simulation, when $L \gtrsim 1$. Under these assumptions, the total cross section, Eq.~(\ref{eq:sigmadef}), can be approximated as
\begin{equation}
\sigma = \frac{4\pi}{\mu^2 v_{\rm rel}^2} (1+L)^2~. \label{eq:sigmaapprox}
\end{equation} 
If the phase shifts are exactly maximal, $\sin \delta_\ell =1$, then there is considerable cancellation that occurs in the sum for the transfer cross section (\ref{eq:sigmatr}).  In this limit it is
\be
\sigma_{\rm tr} =  \frac{4\pi}{\mu^2 v_{\rm rel}^2} (1+L)~.
\label{eq:transferapprox}
\ee
However, there is unlikely to be exact cancellation between the different phase shifts so one would expect, for the case with multiple $\ell$-modes contributing, that the answer lies between (\ref{eq:sigmaapprox}) and (\ref{eq:transferapprox}).  Note that in the limit that the potential is turned off, $\alpha \to 0$, $V(r)$ can be treated as a perturbation and the Born approximation is reasonable.  Our approximation is not valid in this limit as we assume that the scattering is maximal for all $\ell \leq L$, whereas in the Born limit $L=0$ and $\delta_0 \sim \alpha m_\chi k/m_\phi^2$, making the cross section velocity independent.

\begin{figure}[t]
\begin{center}
\centerline{
\includegraphics[width=0.45\columnwidth]{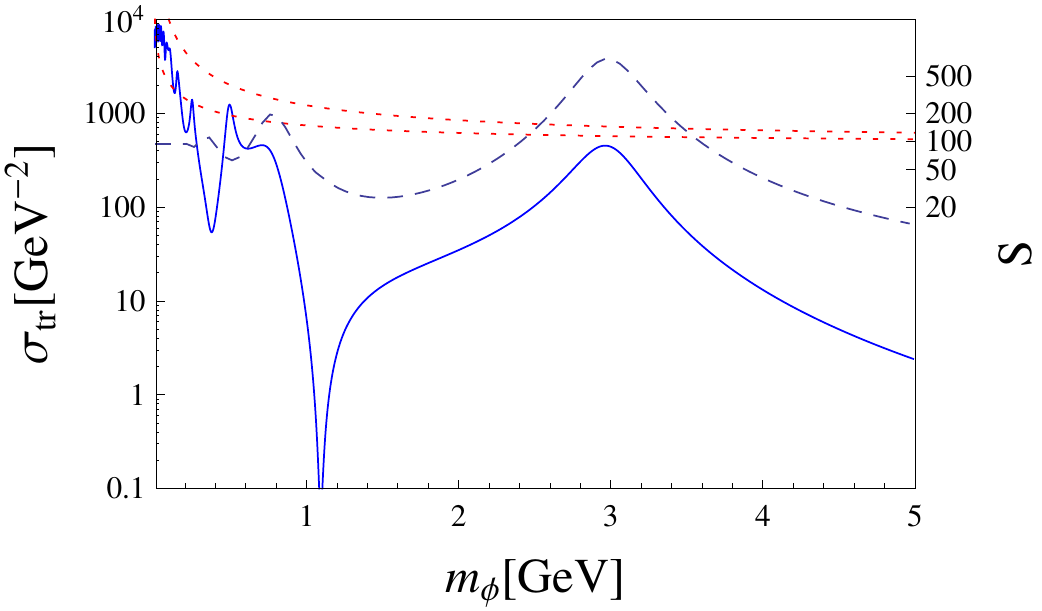}\hspace{0.07\columnwidth}
\includegraphics[width=0.45\columnwidth]{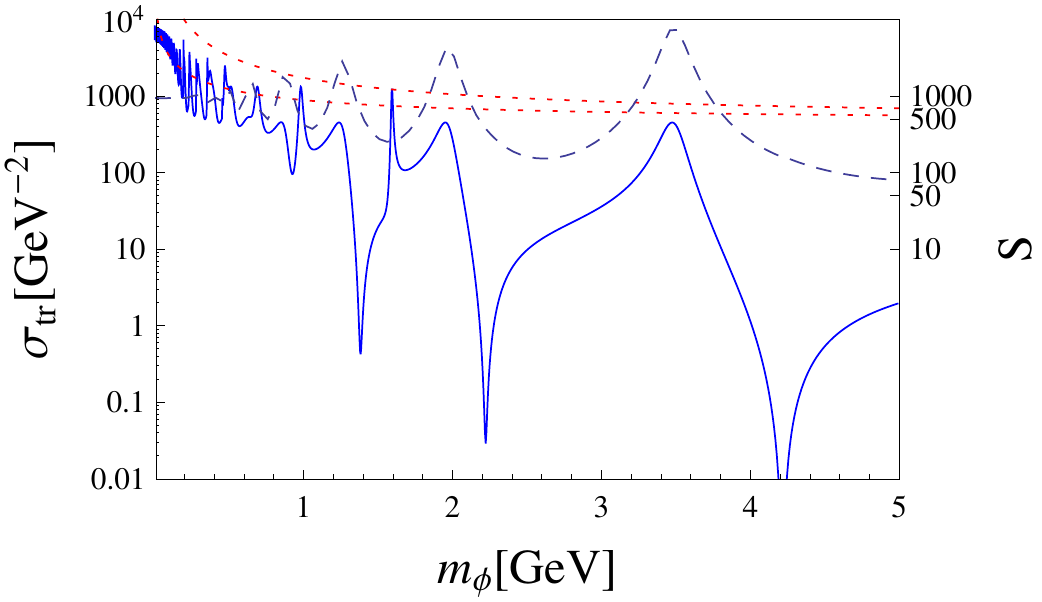}
}
\caption{For $m_\chi=500\ \gev$, and $\alpha = 0.01$ (left-hand plot) and $\alpha=0.1$ (right-hand plot) we show the transfer cross section, the numerical results (blue solid line) and our approximate formulae Eqs.~(\ref{eq:sigmaapprox}) and (\ref{eq:transferapprox}) (upper and lower red dotted lines), as well as the Sommerfeld enhancement (blue dashed line) in the annihilation cross section.  We have assumed that the DM is colliding head-on with speed 100 km/s.}
\label{fig:nointplots}
\end{center}
\end{figure}

\begin{figure}[htbp]
\begin{center}
\centerline{
\includegraphics[width=0.45\columnwidth]{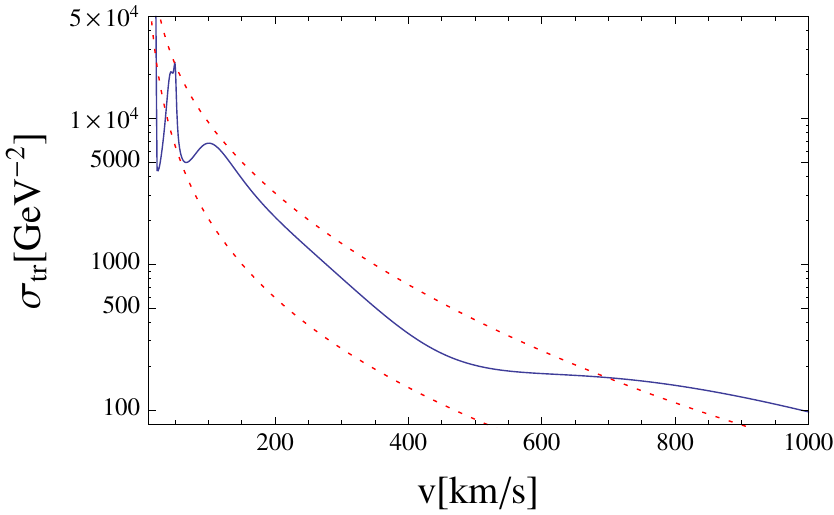}\hspace{0.07\columnwidth}
\includegraphics[width=0.45\columnwidth]{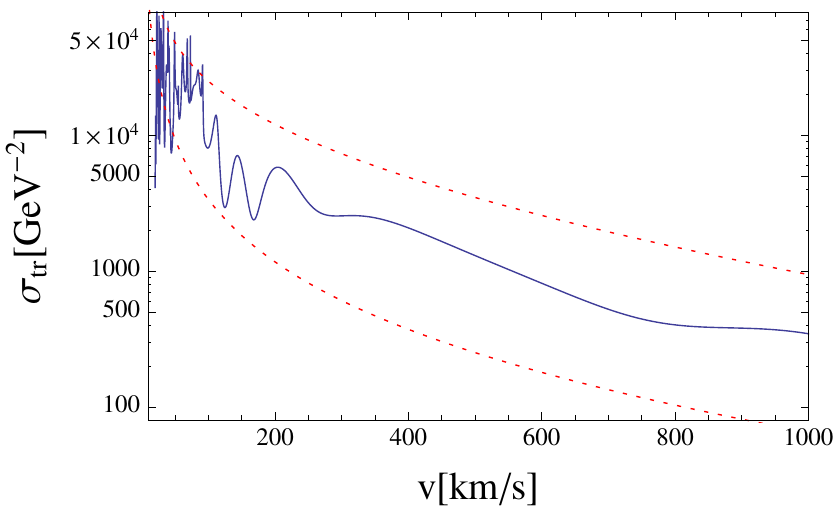}
}
\caption{For $m_\chi = 500\ \gev$, $m_\phi=100\ \Mev$ and $\alpha = 0.01$ (left-hand plot) and $\alpha=0.1$ (right-hand plot) we plot the transfer cross section for two DM particle colliding head on at speed $v$.  The result of the numerical calculation, summing the first five $\ell$-modes, is shown in blue (solid) and the upper and lower red (dotted) curves uses the approximate cross section Eqs.~(\ref{eq:sigmaapprox}) and (\ref{eq:transferapprox}) described in the text.}
\label{fig:vdependence}
\end{center}
\end{figure}

We explicitly calculate the transfer cross section by summing the phase shifts for the low-lying $\ell$-modes.  The region of parameter space for which there are observational constraints has velocities in the range $10~\mathrm{km/s} \ltap v\ltap 1000$~km/s, and we limit ourselves to mediator masses above $\sim 10\ \mev$.  In Fig.~\ref{fig:nointplots} we present the resulting $\sigma_{\rm tr}$, for DM of fixed speed of 100 km/s and two choices of $\alpha$ ($\alpha=0.01$ and $\alpha=0.1$). For $m_\phi< 500$ MeV we sum the contributions from $\ell\le 5$ and for larger $m_\phi$ only the first two modes are included. In Fig.~\ref{fig:vdependence} we illustrate the dependence of $\sigma_{\rm tr}$ upon $v$ for fixed mediator mass, for the same two values of $\alpha$. Although we have summed up several $\ell$ modes, for all but very low mediator masses the cross section is dominated by the $s$-wave.  As anticipated, there are points in parameter space with nearby bound states, causing a resonance in the scattering which also results in an increase in the annihilation rate (dashed blue line in Fig.~\ref{fig:nointplots}). 

It can also be seen in Fig.~\ref{fig:nointplots} that our simple approximation (dotted red lines) does a good job of capturing the correct behaviour at low mediator mass, but at higher mediator masses it significantly overestimates the result away from the positions in parameter space where there is a resonance. This is as expected, since at a resonance $\sin \delta_\ell \to 1$, which is the saturation assumption made in the derivation of the approximate formula Eq.~(\ref{eq:transferapprox}).

This approximation is also useful since solving the Schr\"odinger equation (\ref{eq:rescaledschrodinger}) numerically for a large number of $\ell$-modes is time consuming.  As mentioned earlier, if the coupling, $\alpha$ is weak or the velocity of the DM large then the potential is a small effect and the 
Schr\"odinger equation can be solved perturbatively, the Born approximation.  However, for the low relative velocities of interest here the Born approximation is not valid over much of the parameter space.  Alternatively, one can numerically solve the classical motion of a particle in the same Yukawa potential.  For a large range of parameters these numerical results are well fit by a simple analytic expression~\cite{PhysRevLett.90.225002,Feng:2009hw}.  While the classical result will miss the existence of resonances present in the quantum mechanical solution, since classically there can be no tunneling, the two approaches agree away from these special, but not negligible, regions of parameter space.

\section{Observational Bounds}

Self-interactions transfer energy between DM particles and thus tend to drive DM halos into a spherical isothermal configuration, with an approximately Maxwellian phase space density. The observation of non-spherical halos and halos with large phase-space densities (dwarf spheroidals and the cores of galaxy clusters) place constraints on the size of the self-interactions. A weaker bound has also been derived from the Bullet Cluster. These bounds are summarized in Table I, along with the characteristic velocities of the relevant systems. We review them individually here.\footnote{The unit of choice for $\sigma/m_\chi$ in $N$-body simulations is the cm$^2$/g. We adopt this convention for the remainder of the paper. For convenience, note that $1$~GeV$^{-3} \sim 2\times 10^{-4}$~cm$^2$/g. For an overview of the constraints, see \cite{Taoso:2007qk}.}

The Bullet cluster bounds \cite{Clowe:2006eq,Markevitch:2003at,Randall:2007ph} are weak, but conceptually simple. The Bullet cluster system consists of two galaxy clusters which have collided. The baryonic gas has been slowed by the collision, while gravitational lensing demonstrates that the two DM halos have flowed freely through each other. From this, a robust bound of $\sigma/m_\chi \leq 1.25$~cm$^2$/g has been placed on the self-scattering; however, by making additional assumptions about the initial states of the two clusters one can strengthen this bound to $\leq 0.7$~cm$^2$/g \cite{Randall:2007ph}.

Smaller DM halos must be colder than larger ones, as fast moving DM particles can escape from a smaller gravitational well. If a smaller subhalo is in orbit around a larger halo ({\it i.e.}~dwarf galaxies around a spiral galaxy or a spiral galaxies around a galactic cluster) then large cross sections allow the efficient transfer of heat from the hot DM in the large halo to the colder DM in the subhalo. This causes the subhalo to dissipate, either through spallation or evaporation (depending on the size of the scattering cross section)~\cite{Wandelt:2000ad}. In order that galaxies such as our own do not evaporate too quickly in the hot DM of the host cluster, the range $0.3~\mbox{cm$^2$/g} \leq \sigma/m_\chi \leq 10^4~\mbox{cm$^2$/g}$ may be excluded \cite{Gnedin:2000ea}. In addition, if the cross section is allowed to depend on powers of velocity, {\it i.e.}~$\sigma/m_\chi = \sigma_0/m_\chi (v/v_0)^{-2\delta}$, then the rate of evaporation of dwarf galaxies rules out the range $1.1 < \delta < 1.8$ for $\sigma_0/m_\chi = 0.1$~cm$^2$/g~\cite{Gnedin:2000ea}.  For such bounds the characteristic velocity is that of the larger, hotter halo, rather than the smaller and colder one. 

In addition to causing the dissipation of smaller halos, self-scattering also causes the cores of DM halos to become more spherical, larger, and less cuspy, as the increased rate of energy transfer allows smoothing over the central region. As the cores of some galaxy clusters are known to be elliptical \cite{MiraldaEscude:2000qt,Arabadjis:2001cm,Buote:2002wd}, a limit of $\sigma/m_\chi < 0.02$~cm$^2$/g can be derived from these systems \cite{MiraldaEscude:2000qt}.  
Previously, the lack of cusps in the cores of dwarf galaxies, combined with a paucity of dwarf subhalos relative to the number of large galaxies in the Local Group (the `missing satellite problem'), 
was taken as evidence in favor of collisional dark matter \cite{Spergel:1999mh}. The range of $0.5~\mbox{cm$^2$/g} \lesssim \sigma/m_\chi \lesssim 5$~cm$^2$/g was claimed to put $N$-body simulations in better correspondence with observation \cite{Spergel:1999mh,Dave:2000ar,Yoshida:2000uw,Wandelt:2000ad,Colin:2002nk,Ahn:2002vx}.  
However, it since has been shown that tidal stripping \cite{Kravtsov:2004cm} or photoionization effects \cite{Bullock:2000wn} can bring simulation in line with observation. In this case, no additional scattering is necessary to explain the structure of dwarf galaxies, and indeed, we can place a bound of $\sigma/m_\chi \lesssim 0.06$~cm$^2$/g by requiring that the $N$-body predications do not differ from collisionless DM \cite{Hannestad:2000bs}. To be conservative, we take this dwarf galaxy bound to be $0.1$~cm$^2$/g.

Studies of the thermodynamics of galaxies indicate that if the average time between collisions is less than Hubble time, then the cusps of dark matter cores would be flatter and larger than what we observe \cite{Firmani:2000qe,Hui:2001wy,Colin:2002nk}. In order that this not occur, the cross section is be bounded by \cite{Firmani:2000qe}
\begin{equation}
\frac{\sigma}{m_\chi} \lesssim 0.2 ~\mbox{cm$^2$/g} \left(\frac{0.02 M_{\odot}\mbox{pc}^3}{\rho} \right)\left(\frac{100~\mbox{km/s}}{v_0}\right). \label{eq:collision}
\end{equation}
Here, $\rho$ is the DM density of the system. The DM systems considered in Ref.~\cite{Firmani:2000qe} have velocities and densities such the cross section is limited to be $\lesssim 0.01-0.6$~cm$^2$/g. 

Finally, the rate of growth of super-massive black holes (such as the one in the core of our own Galaxy) place a limit on the scattering of dark matter \cite{Ostriker:1999ee,Hennawi:2001be}. If dark matter is collisional, then supermassive black holes would grow faster than in the collisionless case; rapidly reaching the point where the size of the accretion disk approaches themean free path of the DM. If we take the cross section to be velocity dependent $\sigma = \sigma_0 (v/v_0)^a$, then the bound becomes \cite{Ostriker:1999ee}
\begin{equation}
\sigma_0/m_\chi\left(\frac{v_0}{100~\mbox{km/s}}\right)^a \lesssim 0.02~\mbox{cm$^2$/g}. \label{eq:blackhole}
\end{equation}

It should be noted that, with a few exceptions outlined above, most simulations to date model DM self-interactions as a classical hard-sphere and thus assume the cross section is velocity independent, {\it i.e.}~$d\sigma/d\Omega = b^2$.  By allowing the scattering cross-section to depend on inverse powers of $v$, the more restrictive bounds from clusters could in principle be avoided, while still allowing large cross sections at low velocities that could  alter the structure of dwarf galaxies from the standard dark matter prediction. 

However, as the majority of the constraints are not derived from analyses that allowed for velocity-dependent cross sections, it is not completely clear whether we can apply these bounds to our work. One should bear in mind that the effects of large scattering are somewhat difficult to anticipate; for example, we expect larger cross sections to increase energy transfer rates, however if the mean free path becomes much smaller than the relevant length scales in the system then the medium becomes optically thick and transfer rates actually decrease.

Since we cannot repeat a full $N$-body simulation with the velocity dependent scattering cross section given by Eq.~(\ref{eq:diffcrosssection}), we instead calculate the transfer cross section, averaged over the DM velocity distribution in the appropriate system, and compare to the corresponding bound on $\sigma/m_\chi$. The transfer cross section, rather than the unweighted cross section, is the parameter of interest since it measures the rate at which energy is transferred in the system. As outlined above, this is the quantity being constrained in the various limits, either directly or indirectly.
In particular we calculate
\be
\vev{\sigma_{\rm tr}} = \int d^3 \vec{v}_1\, d^3 \vec{v}_2\, d\Omega\  f(\vec{v}_1)f(\vec{v}_2) (1-\cos\theta) \frac{d\sigma}{d\Omega} ~,
\label{eq:avtransfercs}
\ee 
where in the galaxy's frame the WIMPs have velocity distribution 
\be
f(v)=\frac{1}{(\pi v_0^2)^{3/2}}e^{-v^2/v_0^2}~.
\label{eq:veldis}
\ee
Though we have, for completeness, listed a number of bounds in Table~\ref{tab:bounds}, we shall show that the one of most interest is the dwarf galaxy bound, as it comes from the system with the lowest dispersion velocity. For comparison, we shall also consider in detail the bounds placed by elliptic clusters, as those represent the tightest bounds on a system with high velocity DM.

\begin{table}[t]
\begin{center}
\begin{tabular}{|c|c|c|c|}
\hline System & $v_0$[km/s] & $\sigma/m_\chi\,\left[\mathrm{cm}^2/\mathrm{g}\right]$ & References\\
\hline
Bullet Cluster & 1000 & 1.25 & \cite{Clowe:2006eq,Randall:2007ph} \\
Galactic Evaporation & 1000 & 0.3 & \cite{Gnedin:2000ea} \\ 
Elliptic Cluster & 1000 & 0.02 & \cite{MiraldaEscude:2000qt} \\ 
Dwarf Evaporation & 100 & 0.1$^\star$ & \cite{Gnedin:2000ea} \\ 
Black Hole & 100 & 0.02$^\star$ &  \cite{Ostriker:1999ee} \\ 
Mean Free Path & $44-2400$ & $0.01-0.6$ & \cite{Firmani:2000qe} \\
Dwarf Galaxies & 10 & 0.1 & \cite{Hannestad:2000bs} \\ \hline
\end{tabular}
\end{center}
\caption{The systems we consider and the observational bound they place on DM self-scattering cross section. Entries marked with an asterisk $^\star$ are velocity dependent bounds. For more details, see text.}
\label{tab:bounds}
\end{table}%

\begin{figure}[ht]
\centerline{
\includegraphics[width=0.45\columnwidth]{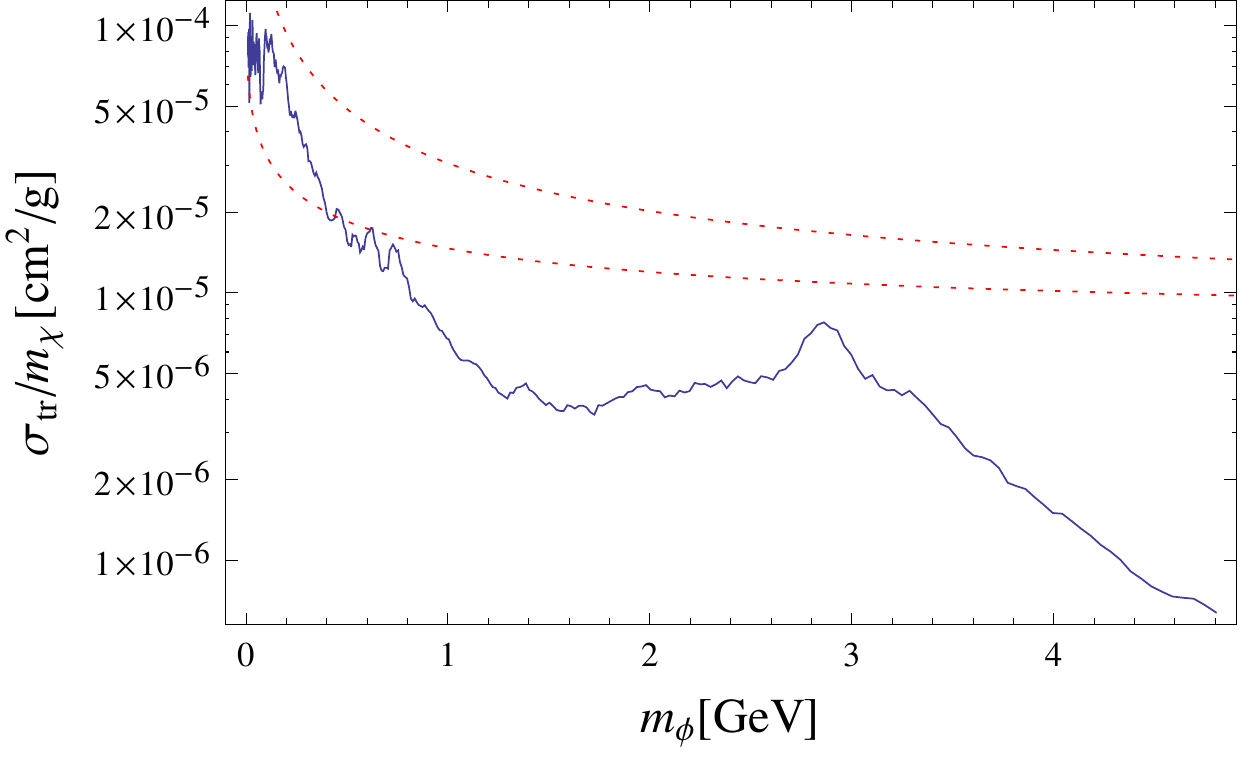}\hspace{0.07\columnwidth}
\includegraphics[width=0.45\columnwidth]{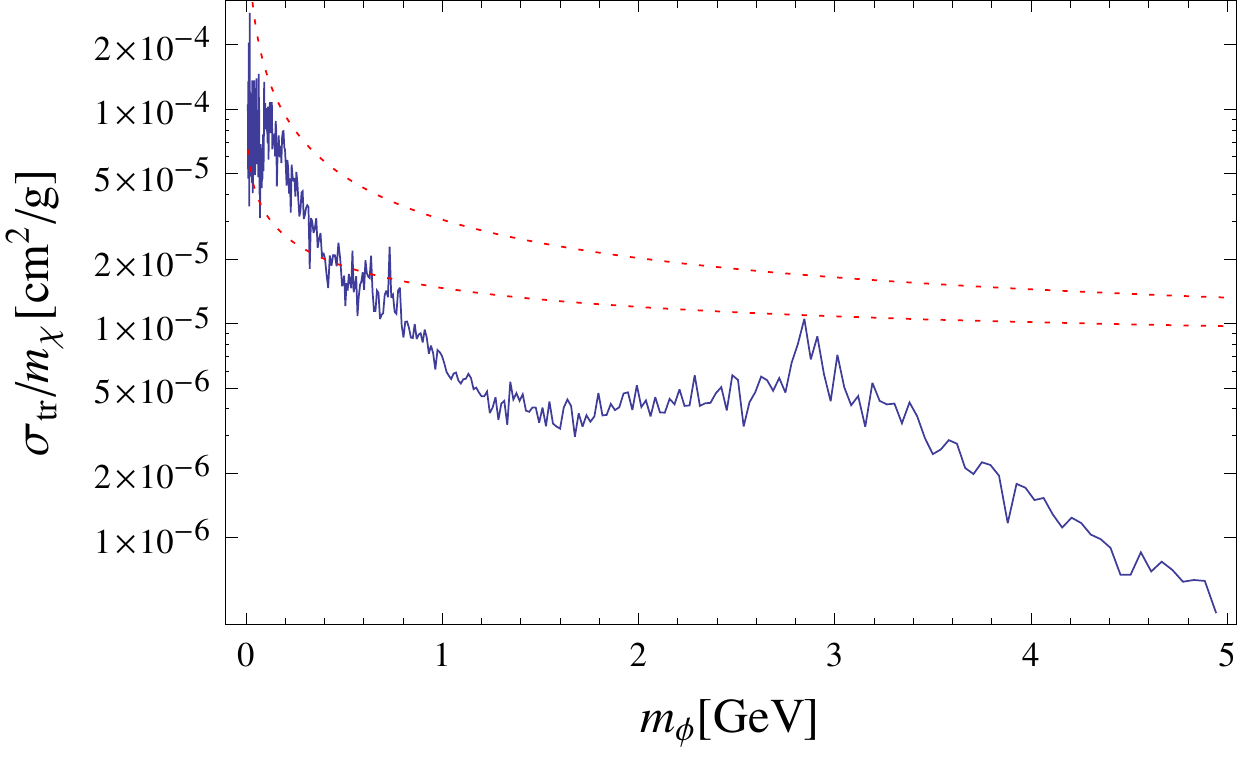}
}
\caption{$\langle \sigma_{\rm tr} \rangle /m_\chi$ as a function of $m_\phi$, assuming that $m_\chi=500$~GeV, and $\alpha = 0.01$ (left) and $\alpha = 0.1$ (right). A thermal velocity distribution Eq.~(\ref{eq:veldis}) with dispersion $v_0 = 1000~\mbox{km/s}=3.3\times 10^{-3}c$, characteristic of galaxy clusters, was used. Contributions from modes up to $\ell = 5$ are included in the exact numerical cross section for $m_\phi < 0.2$~GeV, while only $\ell \leq 1$ are included above this mass. The approximate solutions from Eqs.~(\ref{eq:sigmaapprox}) and (\ref{eq:transferapprox}) are also shown (dashed red lines). \label{fig:cluster}}
\end{figure}

In Fig.~(\ref{fig:cluster}) we show $\langle \sigma_{\rm tr} \rangle/m_\chi$ as a function of $m_\phi$, assuming a Maxwellian distribution with characteristic velocity $v_0 = 1000$~km/s, which is approximately the value found in galaxy clusters. For values of $m_\phi$ greater than $0.5$~GeV we include only $\ell$ modes of zero and one, while for $m_\phi < 0.5$~GeV, we include $\ell \leq 5$. As can be seen in Fig.~\ref{fig:lmax}, for our choice of $\alpha$, $m_\phi$ and $v$, this is an acceptable trade-off between computational speed and accuracy. We also display the approximate solutions for the cross section and transfer cross section, as given by Eqs.~(\ref{eq:sigmaapprox}) and (\ref{eq:transferapprox}), again integrating over a Maxwellian distribution for both incoming particles (the upper line is the approximate cross section, while the lower is $\sigma_{\rm tr}$).  We can clearly see that for systems with velocity distributions centered around 1000~km/s no bounds on MeV-scale dark forces can be placed.

\begin{figure}[ht]
\centerline{
\includegraphics[width=0.45\columnwidth]{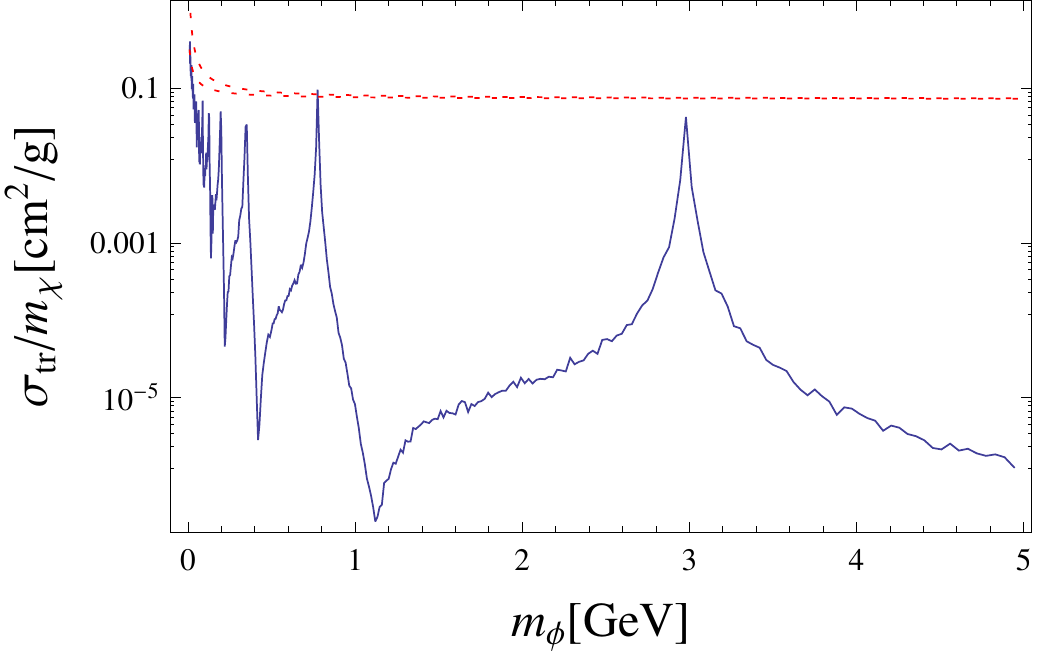}\hspace{0.07\columnwidth}
\includegraphics[width=0.45\columnwidth]{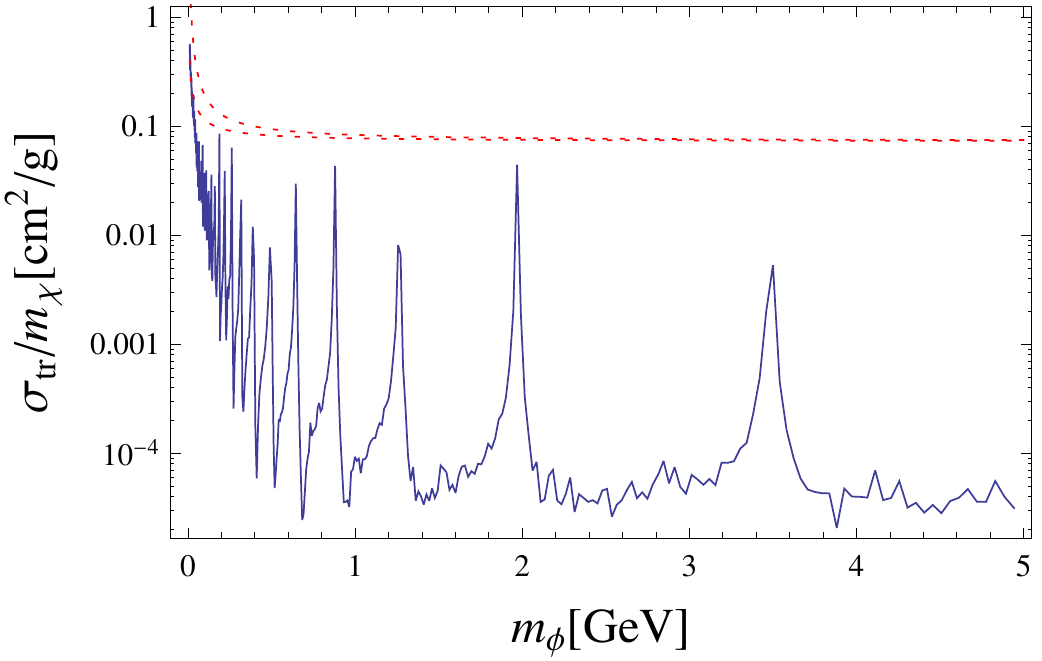}
}
\centerline{
\includegraphics[width=0.45\columnwidth]{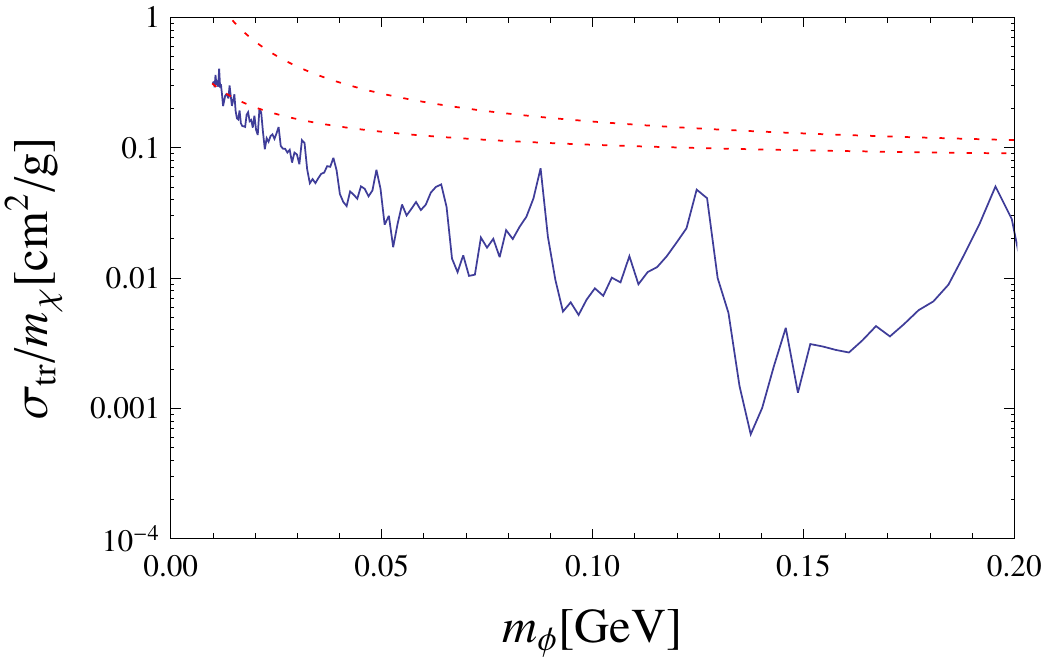}\hspace{0.07\columnwidth}
\includegraphics[width=0.45\columnwidth]{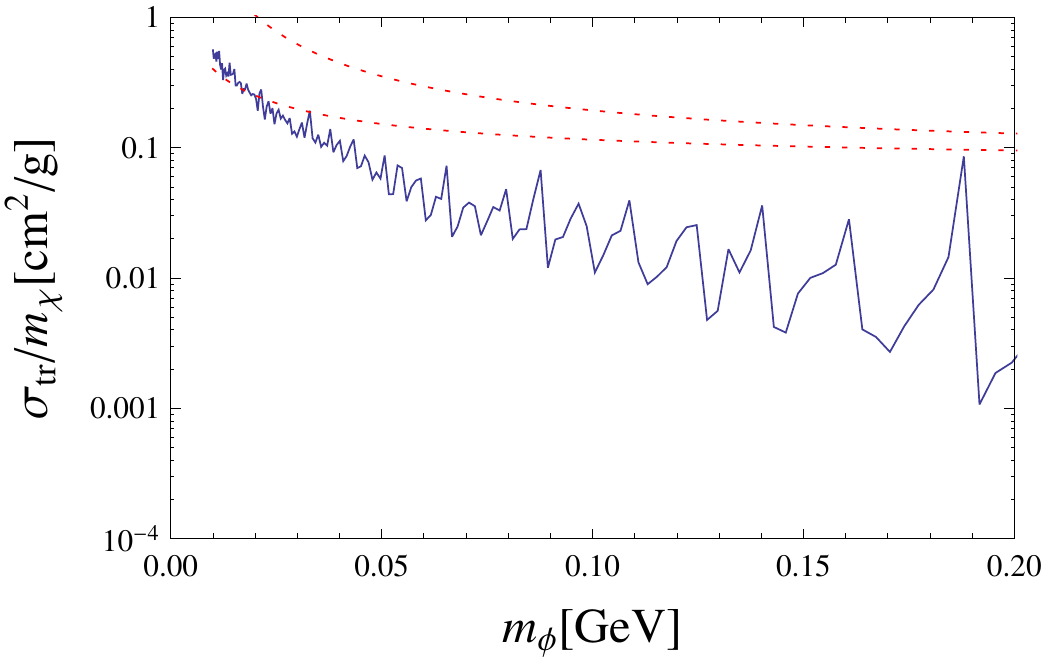}
}

\caption{$\langle \sigma_{\rm tr} \rangle /m_\chi$ as a function of $m_\phi$, assuming that $m_\chi=500$~GeV, and $\alpha = 0.01$ (left) and $\alpha = 0.1$ (right). A thermal velocity distribution Eq.~(\ref{eq:veldis}) with dispersion $v_0 = 10~\mbox{km/s}=3.3\times 10^{-5}c$, characteristic of a dwarf galaxy, was used. Contributions from modes up to $\ell = 5$ are included in the exact numerical cross section. The approximate solutions from Eqs.~(\ref{eq:sigmaapprox}) and (\ref{eq:transferapprox}) are also shown (red dashed lines). Regions of low $m_\phi$ from the upper plots are shown in more detail in the lower plots. The limit from dwarf galaxy structure is $0.1$~cm$^2$/g. \label{fig:dwarf}}
\end{figure}

However, dwarf galaxies, with velocity dispersions of $\sim 10$~km/s \cite{Strigari:2006rd}, provide a non-trivial constraint. In Fig.~\ref{fig:dwarf} we show the velocity-averaged $\langle \sigma_{\rm tr}\rangle/m_\chi$ as a function of $m_\phi$, this time for a dwarf galaxy-appropriate value of $v_0 = 10$~km/s. Again, both the exact numerical solution (with $\ell \leq 5$ for all values of $m_\phi$) and the approximate solutions are shown. Taking the upper bound on $\sigma/m_\chi$ to be the $0.1$~cm$^2$/g derived from dwarf galaxies, we can place a bound requiring
\begin{equation}
m_\phi \gtrsim 40~\mbox{MeV} \label{eq:bound1} 
\end{equation}
for the larger value of $\alpha$ considered and slightly weaker ($m_\phi\gtap 30$ MeV) for smaller $\alpha$.  Although clusters present a tighter bound on the scattering cross section, the characteristic velocity in these systems is far higher (Table~\ref{tab:bounds}) and the stronger constraint comes from dwarf galaxies.  

A full simulation for the case of velocity dependent cross sections, as expected in models with a Sommerfeld enhancement, would improve on our estimate and we advocate strongly for it to be carried out.  

\section{Conclusions}

An appealing explanation of the recent cosmic ray excesses is in terms of DM annihilation in our local galactic neighborhood, with the necessary increase in annihilation cross section (above that for a typical thermal WIMP) being due to a new dark force.  The exchange of this dark force carrier $\phi$ -- either a scalar or vector boson -- leads to an attractive force between dark matter particles and consequently an increase at low velocities in their annihilation cross section: the Sommerfeld enhancement.  To get a large enough enhancement the force must be sufficiently long-range and the force carrier is typically taken to be sub-GeV in mass. This is done for two reasons.  The first is that the enhancement approximately scales as $\sim \alpha m_\chi/m_\phi$, although it should be noted that for Yukawa potentials there are nearly massless bound states that lead to a far greater enhancement and a scan over parameter space shows that the density of these bound states is relatively high.  The second is that the cosmic ray excesses are in leptonic channels and one way this might occur is if the DM annihilates to $\phi$ particles which in turn decay to SM states.  If the force carrier is sub-GeV in mass scale its decay products will be leptons.

Solving the Schr\"odinger equation for the case of a Yukawa potential between the DM allows us to go beyond the Born approximation, resumming the effects of multiple light boson exchange, and to capture the effects of resonances in the potential.  We have shown that the physics which causes the large enhancement in annihilation necessarily implies an increase in the self-scattering cross section.  Requiring that this scattering enhancement does not cause significant energy transfer inside dark matter halos allows us to constrain the possible mass of the mediator $m_\phi$, independent of the details of how it couples to the SM.  There are various bounds on the self-scattering cross section coming from multiple systems.  Approximately speaking, they amount to requiring less than one DM scattering per Hubble time, but by comparing $N$-body simulations to observations this approximate bound can be improved upon in certain cases.  

Unfortunately many of these systems have very different DM velocity distributions and this hinders easy translation from a bound on one system to another, as most $N$-body simulations assume a velocity independent hard sphere cross section for DM scattering.  We advocate strongly for $N$-body simulations to be carried out which include the velocity dependence appropriate to the light mediator in the DM cross section.  To allow comparison between these bounds and our velocity dependent scattering we calculate the transfer cross section averaged over the DM velocity distribution in two systems: a galaxy cluster and a dwarf galaxy.  We find that galaxy clusters offer no interesting bound whereas dwarf galaxies require that $m_\phi \gtap 40\ \Mev$, with some sensitivity to the DM-mediator coupling.  For mediators heavier than this bound there is still considerable Sommerfeld enhancement, both close to and away from resonances.  This bound may well be improved upon by detailed simulations and or as further observations become available.

Direct searches for a new $U(1)$ gauge boson which would play the role of  $\phi$ have been suggested in $e^-e^+$ collisions, and will be sensitive to $m_\phi$ of ${\cal O}(1~\mbox{GeV})$ \cite{Bjorken:2009mm}. Current bounds already exist for a range of $m_\phi$ and photon-$\phi$ mixing from  $\Upsilon(3S)$ decay at BaBar \cite{:2009cp}, measurements of $e$ and $\mu$ anomalous magnetic moments \cite{Pospelov:2008zw}, the beam dump experiments at SLAC \cite{Bjorken:1988as,Riordan:1987aw} and Fermilab \cite{Bross:1989mp}, and dark photon searches at the Tevatron~\cite{Abazov:2009hn}. However, these experiments, while probing the same mass range as Eq.~(\ref{eq:bound1}) depend not only on $m_\phi$, but also on its couplings to the SM. Our result is independent of the mixing with the photon, and in fact holds regardless of whether $\phi$ is a $U(1)$ gauge boson, a gauge boson of a non-abelian gauge group, or a scalar field.
In that sense, the result is very robust.

In conclusion, we have found that, for an experimentally interesting range of parameters, the proposed new `dark forces' of interest in explaining results from both direct and indirect DM searches would have measurable impact on the structure of dwarf galaxy halos. As the enhancement of dark matter self-scattering does not depend on the details of the attractive force carrier $\phi$ beyond its mass and self-coupling, these bounds are applicable across a wide range of possible scenarios. Indeed, the increase in the scattering cross section could be present even in models where DM self-annihilation was forbidden by some symmetry. However, at this point, more work is necessary in order to be confident that the bounds on $\sigma/m_\chi$, derived from $N$-body simulation assuming a velocity independent cross section, can be applied in to situations where $\sigma_{\rm tr}$ depends on inverse powers of $v$. From the calculations provided in this paper, we believe that investigating this possibility is a worthwhile task, as it does seem that the effects are large enough to offer interesting constraints.

\section*{Acknowledgements}
We thank Manoj Kaplinghat, and Hai-Bo Yu for informative discussions.
We would like to thank Neal Weiner for helpful discussions and comments on a draft of this work, and the Galileo Galilei Institute for Theoretical Physics for hospitality while part of this work was completed.  MRB would like to thank Sean Carroll, Dan Hooper, and Mark Wise for useful discussion and advice. He also thanks the Aspen Center for Physics for providing a simulating environment for collaboration and writing. PJF would like to thank Erich Poppitz for collaboration during the early stages of this work.  
MRB is supported by the Department of Energy, under grant DE-FG03-92-ER40701.   Fermilab is operated by Fermi Research Alliance, LLC under contract no. DE-AC02-07CH11359 with the United States Department of Energy

\emph{Note Added:} During the completion of this work we became aware of a paper that addresses similar issues~\cite{Feng:2009hw}.  Though they consider a different astrophysical system we reach similar conclusions.


\bibliography{scattering}
\bibliographystyle{apsrev}

\end{document}